\documentclass[aps,prd,showkeys,superscriptaddress,twocolumn,nofootinbib,floatfix]{revtex4-2}

\usepackage{subfiles}

\usepackage{amsmath}
\usepackage{amssymb}
\usepackage{amsthm}
\usepackage{mathrsfs}
\usepackage{graphicx}
\usepackage{epstopdf}
\usepackage{fancyhdr}
\usepackage{array}
\usepackage[all]{xy}
\usepackage{eufrak}
\usepackage{euscript}
\usepackage{enumerate}
\usepackage{slashed}
\usepackage{hyperref}
\usepackage{subfigure} 
\usepackage{epstopdf} 

\hypersetup{pdftex,colorlinks=true,linkcolor=blue,citecolor=blue,menucolor=black,urlcolor=blue,filecolor=blue}

\begin{document}

\title{Holographic Bjorken flow of a hot and dense fluid in the vicinity of a critical point}

\author{Renato Critelli}
\email{renato.critelli@usp.br}
\affiliation{Instituto de F\'{i}sica, Universidade de S\~{a}o Paulo, Rua do Mat\~{a}o, 1371, Butant\~{a}, CEP 05508-090, S\~{a}o Paulo, S\~{a}o Paulo, Brazil}

\author{Romulo Rougemont}
\email{romulo.pereira@uerj.br}
\affiliation{Departamento de F\'{i}sica Te\'{o}rica, Universidade do Estado do Rio de Janeiro,
Rua S\~{a}o Francisco Xavier 524, 20550-013, Maracan\~{a}, Rio de Janeiro, Rio de Janeiro, Brazil}

\author{Jorge Noronha}
\email{noronha@if.usp.br}
\affiliation{Instituto de F\'{i}sica, Universidade de S\~{a}o Paulo, Rua do Mat\~{a}o, 1371, Butant\~{a}, CEP 05508-090, S\~{a}o Paulo, S\~{a}o Paulo, Brazil}

\begin{abstract}
In this paper we use the gauge/gravity duality to perform the first systematic study of the onset of hydrodynamic behavior in a hot and dense far-from-equilibrium strongly coupled relativistic fluid with a critical point. By employing a top-down holographic construction that stems from string theory, we numerically obtain the full nonlinear evolution of the far-from-equilibrium system undergoing a Bjorken expansion and address the following question: how does hydrodynamic behavior emerge in the vicinity of a critical point in the phase diagram? For the top-down holographic system analyzed in the present work, we find that the approach to hydrodynamics is strongly affected by the presence of the critical point: the closer the ratio between the chemical potential and the temperature is to its critical value, the longer it takes for the system to be well  described by the equations of viscous hydrodynamics.
\end{abstract}



\maketitle

\section{Introduction}
\label{sec:intro}

Quantum chromodynamics (QCD) \cite{Gross:1973id,Politzer:1973fx} is the fundamental theory of nature that describes the interactions between quarks and gluons, accounting for the vast majority of the mass of ordinary baryonic matter in the Universe \cite{Weinberg:2008zzc}. QCD is also responsible for gluing protons and neutrons together inside atoms allowing for the stability of nuclear matter and a long-standing challenge in the field has been the determination of its equilibrium properties as a function of temperature $T$ and baryon chemical potential $\mu_B$ \cite{Philipsen:2012nu}. 

The QCD phase diagram has been the focus of many experimental, observational, and theoretical efforts through the years. In the experimental front, the main tool used to investigate hot and dense QCD matter is heavy ion collisions \cite{Arsene:2004fa,Adcox:2004mh,Back:2004je,Adams:2005dq,Aad:2013xma}, which produce a strongly coupled fluid made of deconfined but highly correlated quarks and gluons, the so-called quark-gluon plasma (QGP) \cite{Gyulassy:2004zy,Heinz:2013th,Shuryak:2014zxa}. In the observational front, the core of compact neutron stars may comprise new phases of quark matter expected to exist when $\mu_B/T\gg 1$, such as color superconductivity \cite{Alford:2007xm}. While the equation of state of such high baryon dense matter is currently unknown, powerful constraints may be derived at low and asymptotically high densities \cite{Kurkela:2014vha,Fraga:2015xha} which together with observational data on neutron stars \cite{Demorest:2010bx,Antoniadis:2013pzd}, including the very recent observation of gravitational waves emitted by a neutron star merger \cite{TheLIGOScientific:2017qsa,Annala:2017llu}, provide a much better understanding of QCD under those extreme conditions.

In the theoretical front, first principles lattice QCD simulations can successfully describe many equilibrium properties of the QGP up to $\mu_B/T\sim 2$ \cite{Bazavov:2017dus,Ratti:2018ksb} including the crossover nature of the QCD phase transition at zero chemical potential \cite{Aoki:2006we,Borsanyi:2016ksw}. However, even lattice QCD simulations suffer from severe limitations, as they are currently unable to deal with larger values of baryon chemical potential (due to the fermion sign problem \cite{Philipsen:2012nu}) and also real time out-of-equilibrium phenomena \cite{Asakawa:2000tr}, which are of fundamental importance for the spacetime evolution of the fireball produced in heavy ion collisions and also for the cold and dense matter core of neutron stars.

In particular, a crucial question for the beam energy scan (BES) program conducted at RHIC \cite{Aggarwal:2010cw}, the future fixed target experiments at RHIC \cite{Meehan:2016qon,Meehan:2017cum}, the compressed baryonic matter (CBM) experiment at FAIR \cite{Staszel:2010zza,Ablyazimov:2017guv}, and also the upcoming experiments at NICA \cite{NICA}, is the location of the QCD critical end point (CEP), which would mark the end of a first order phase transition line expected to exist in the QCD phase diagram at nonzero baryon density \cite{Stephanov:1998dy,Stephanov:1999zu,Rischke:2003mt,Stephanov:2011pb}.

If the QCD CEP is within the reach of low energy heavy ion collisions, as predicted by a recent phenomenological holographic model \cite{Critelli:2017oub} which is in quantitative agreement with state-of-the-art lattice QCD results \cite{Bazavov:2017dus,Bellwied:2015lba}, the rapidly expanding medium formed in these collisions may generally pass close to this point when it is still far from thermal equilibrium being perhaps not even close to a hydrodynamical regime. Therefore, the investigation of how the presence of a critical point affects the far-from-equilibrium dynamics of the expanding medium, and its subsequent approach to hydrodynamics, is of fundamental importance. In the absence of first principle QCD calculations of the full dynamical evolution of the system, one must resort to effective models. Holographic shockwave-like simulations of a strongly coupled plasma at finite chemical potential (though without critical phenomena) were investigated in \cite{Casalderrey-Solana:2016xfq} while the spinodal instability in a first-order phase transition at strong coupling was studied in \cite{Attems:2017ezz}. A way to extend hydrodynamics through the inclusion of parametric slowing down and critical fluctuations induced by a critical point has been recently proposed in \cite{Stephanov:2017ghc}. Also, recently out-of-equilibrium effects have been shown to produce significant impact on the critical behavior of the cumulants of fluctuations of conserved charges, which has been investigated using the Fokker-Planck equation \cite{Mukherjee:2015swa} and also the Kibble-Zurek formalism \cite{Mukherjee:2016kyu}.

In this paper we employ the gauge/gravity correspondence \cite{Maldacena:1997re,Gubser:1998bc,Witten:1998qj,Witten:1998zw} to investigate the effects of a critical point on the hydrodynamization properties of a top-down holographic construction that stems from string theory  known as the 1-R charged black hole (1RCBH) model \cite{Gubser:1998jb,Behrndt:1998jd,Kraus:1998hv,Cai:1998ji,Cvetic:1999ne,Cvetic:1999rb}. This is the holographic dual of a $\mathcal{N}=4$ Supersymmetric Yang-Mills (SYM) plasma at finite temperature and chemical potential featuring a critical point in its phase diagram. Although not directly related to QCD, the 1RCBH model is a very attractive setup for studying far-from-equilibrium phenomena at finite temperature and density in the presence of a critical point. This well-defined holographic construction allows one to investigate many different physical aspects of a strongly coupled relativistic fluid with chemical potential and critical behavior under well controlled theoretical conditions. As a matter of fact, in previous works the thermodynamics and hydrodynamics \cite{DeWolfe:2011ts}, the spectra of quasinormal modes \cite{Finazzo:2016psx}, and the homogeneous isotropization and thermalization dynamics \cite{Critelli:2017euk} of the 1RCBH system have been analyzed in detail (for the calculation of quasinormal modes at criticality in nonconformal settings, see Refs. \cite{Betzios:2017dol,Rougemont:2018ivt}).

In the present work we perform, for the first time in the literature, a \textit{first principles holographic evaluation} of the full far-from-equilibrium evolution of a hot and dense relativistic fluid endowed with a critical point --- the 1RCBH plasma ---, undergoing a boost invariant expansion known as Bjorken flow \cite{Bjorken:1982qr}. The calculations done in the present work are all performed within a single theoretical framework, the gauge/gravity duality, and since the 1RCBH model is a top-down string theory construction, no auxiliary hypotheses or approximations are needed in order to investigate its physical properties. Such a simple and well controlled theoretical setting constitutes the ideal toy model for understanding general aspects of the influence of a critical point on the hydrodynamization dynamics of out-of-equilibrium strongly coupled media. As an example of a general insight coming from the holographic correspondence concerning hydrodynamics, the most robust result obtained from the duality is the nearly perfect fluidity of strongly coupled plasmas, which is encoded in a very small shear viscosity to entropy density ratio, $\eta/s=1/4\pi$ \cite{Kovtun:2004de,Policastro:2001yc,Buchel:2003tz}. This general feature of holographic models is remarkable close to the results of the latest phenomenological simulations describing heavy ion data \cite{Bernhard:2018hnz}.

The properties of a SYM plasma undergoing Bjorken flow at zero chemical potential has been previously investigated using numerical general relativity techniques \cite{Chesler:2009cy,Heller:2011ju,Jankowski:2014lna}, which determined how long it takes for this system to hydrodynamize or, in other words, what is the relevant timescale that characterizes the emergence of hydrodynamic behavior. In the present work, we generalize these studies to the case of the 1RCBH model at finite chemical potential with a critical point. By numerically evolving the full nonlinear partial differential equations of motion of the system in the bulk using a large set of initial conditions that span different values of $\mu/T$, we evaluate the pressure anisotropy, the charge density, the scalar condensate, and the hydrodynamization timescale. The latter describes the time after which a given far-from-equilibrium solution can be reasonably described in terms of the relativistic Navier-Stokes (NS) equations. As the main result of the present work, we show that the onset of hydrodynamics is considerably delayed as the chemical potential is increased towards its critical value.

We use a mostly plus metric signature and natural units $\hbar = c = k_B = 1$.

\section{The holographic model}
\label{sec:HoloDes}

The bulk Einstein-Maxwell-Dilaton (EMD) lagrangian that describes the 1RCBH model is
\begin{align}
2\kappa_5^2\mathcal{L}=R-\frac{(\partial_\mu\phi)^2}{2}-V(\phi) -\frac{f(\phi)(F_{\mu\nu})^2}{4},
\label{eq:EMDaction}
\end{align}
where $\kappa_5^2\equiv 8\pi G_5$ is the five dimensional gravitational constant, $\phi$ is the dilaton field with the scalar potential $V(\phi)= -8e^{\frac{\phi}{\sqrt{6}}} -4 e^{-\sqrt{\frac{2}{3}}\phi}$ (the AdS radius is set to unity), $A_\mu$ is the Maxwell field associated with the chemical potential in the field theory, and $f(\phi)= e^{- 2\sqrt{\frac{2}{3}}\phi}$ is the coupling function between $\phi$ and $A_\mu$.

The Ansatz for the EMD fields in the holographic Bjorken flow, using in-falling Eddington-Finkelstein coordinates, is given by \cite{Chesler:2009cy}
\begin{align}\label{lineElement}
& ds^2 = 2d\tau\left[dr-A(\tau,r) d\tau \right]+\Sigma(\tau,r)^2\left[e^{-2B(\tau,r)}d\xi^2 \right. \notag \\ 
 & \left. + e^{B(\tau,r)}(dx^2+dy^2)\right], \ \ \phi = \phi(\tau,r), \ \ A_\mu dx^\mu = \Phi(\tau,r)d\tau,
\end{align}
where $\tau$ becomes the usual propertime of Bjorken flow at the boundary, $r$ is the radial coordinate of the asymptotically AdS spacetime whose boundary is at $r\rightarrow\infty$, $\xi$ denotes the spacetime rapidity, and $(x,y)$ are the transverse spatial directions to the longitudinal flow direction.

The numerical solutions of the EMD equations accounting for the holographic Bjorken flow in the 1RCBH model are discussed in the Appendix. From these solutions one can extract the energy density ($\varepsilon$), transverse pressure ($p_T$), longitudinal pressure ($p_L$), charge density ($\rho$), and the scalar condensate $\left( \langle \mathcal{O}_{\phi} \rangle \right)$. Also, due to Bjorken symmetry, all of these quantities depend solely on the propertime $\tau$. Because we have a conformal setup, it suffices to look at the energy density and its time derivative to probe the hydrodynamization properties of the system since
\begin{equation}\label{eq:lolol}
\frac{\Delta p}{\varepsilon}\equiv \frac{p_T-p_L}{\varepsilon} = 2+\frac{3}{2}\,\tau\frac{\partial_\tau\varepsilon}{\varepsilon}.
\end{equation}

\section{Viscous relativistic hydrodynamics with a chemical potential}
\label{sec:Hydro_CP}

Now we briefly review the hydrodynamics of a Bjorken expanding fluid with nonzero conserved charge \cite{Erdmenger:2008rm,Banerjee:2008th}, which describes the late time behavior of our numerical gravity simulations. We restrict ourselves to first order hydrodynamics corresponding to the relativistic Navier-Stokes equations \cite{Rezzolla_Zanotti_book}. Higher order derivative corrections to hydrodynamics \cite{Baier:2007ix,Grozdanov:2015kqa} could be implemented once the corresponding transport coefficients (such as the shear viscosity relaxation time) become available for the system under consideration. The improvement in the hydrodynamic description of this system due to the inclusion of higher order gradient corrections, and the subsequent question about the convergence of the series at strong coupling \cite{Heller:2013fn,Buchel:2016cbj}, is left to a future study. 

Due to conformal invariance and the underlying symmetries of Bjorken flow, to first order in the gradient expansion the only contribution to the viscous evolution comes from the shear stress tensor of the system, $\pi^{\mu}_\nu = -\eta\sigma^{\mu}_{\nu}$, where the shear tensor is diagonal with $\sigma^{\mu}_{\nu} \sim 1/\tau$. Since the flow velocity is $u_\mu=(-1,0,0,0)$ and $\pi^\mu_\nu$ is given in terms of the hydrodynamic variables, the NS equations for Bjorken flow with a U(1) global charge reduce to a single equation for the energy density
\begin{equation}
\partial_\tau \varepsilon +\frac{4}{3}\frac{\varepsilon}{\tau}  = \frac{4}{3}\frac{\eta}{\tau^2}, \label{eq:Bjork_Charge} 
\end{equation}
while charge conservation imposes that the charge density associated with the R-charge of the black hole is $\rho(\tau) = \rho_0/\tau$; thus, the charge density is known once one specifies its initial value $\rho_0$.

Using that $\eta/s=1/(4\pi)$ and the well-known thermodynamic relation for conformal theories, $4p =  T s+\mu\rho$, Eq.\ \eqref{eq:Bjork_Charge} reduces to
\begin{equation}\label{eq:eom4}
\partial_\tau \varepsilon +\frac{4}{3}\frac{\varepsilon}{\tau}= \frac{1}{3\pi\tau^2}\left(\frac{4\varepsilon}{3T}-\frac{\mu\rho}{T}\right).
\end{equation}

To proceed, we need the equilibrium equation of state of the 1RCBH model \cite{DeWolfe:2011ts,Finazzo:2016psx,Critelli:2017euk},
\begin{align}
\varepsilon & = \kappa_5^{-2} j(\mu/T)T^4 \Rightarrow T = \kappa_{5}^{1/2}j^{-1/4}\varepsilon^{1/4}, \label{eq:Temp_Bjork_C} \\
\rho & =  \kappa_5^{-2} h(\mu/T)T^3 \Rightarrow \rho = \kappa_5^{-1/2}h j^{-3/4} \varepsilon^{3/4}, \label{eq:ahmoleke}
\end{align}
where,
\begin{align}
j(\mu/T) & = \dfrac{3\pi^4}{32}\left(3-\sqrt{1-\left(\dfrac{\mu/T}{\pi/\sqrt{2}}\right)^2}\right)^3 \notag \\
& \ \ \ \times \left(1+\sqrt{1-\left(\dfrac{\mu/T}{\pi/\sqrt{2}}\right)^2}\right), \\
h(\mu/T) & = \frac{\pi^2}{4}\frac{\mu}{T}\left(3 -\sqrt{1-\left(\dfrac{\mu/T}{\pi/\sqrt{2}}\right)^2}\right)^2 .
\end{align}

Substituting the above results into Eq. \eqref{eq:eom4}, and keeping fixed the dimensionless ratio $\mu/T\equiv x$, it follows that
\begin{equation}\label{eq:eom5}
\partial_\tau \varepsilon +\frac{4}{3}\frac{\varepsilon}{\tau}= \frac{\varepsilon^{3/4}}{3\kappa_{5}^{1/2}\pi\tau^2}\left(\frac{4j^{1/4}}{3}-x \frac{h}{j^{3/4}}\right).
\end{equation}

Using $\hat{\varepsilon}\equiv \kappa_{5}^{2}\varepsilon$ and multiplying by $\tau/\hat{\varepsilon}$
one obtains
\begin{equation}\label{eq:eom7}
\tau \frac{\partial_\tau \hat{\varepsilon}}{\hat{\varepsilon}} = -\frac{4}{3} + \frac{4j-3xh}{9 j^{3/4}\pi w^{(\varepsilon)}},
\end{equation}
where we defined the dimensionless time flow $w^{(\varepsilon)} \equiv \hat{\varepsilon}^{1/4}\tau$. Eq.\ \eqref{eq:eom7} is needed to evaluate the pressure anisotropy in the hydrodynamical regime, according to Eq.\ \eqref{eq:lolol} (which holds also far-from-equilibrium).

Although the definition of $w$ using the energy density seems more natural, as explained in the Appendix, it is possible to use another energy scale $\Lambda$, which is reminiscent from the asymptotic temperature $T_{\textrm{asym}}(\tau)=\Lambda/\left(\Lambda \tau \right)^{1/3}$ \cite{Chesler:2009cy}, to compose another dimensionless time flow, i.e. $w^{(\Lambda)}\equiv T_{\textrm{asym}}(\tau)\tau =\left(\Lambda \tau \right)^{2/3}$. In practice, we use the late time solution of Eq. \eqref{eq:eom7} for the energy density to extract $\Lambda$ for each solution, i.e.
\begin{equation}
\hat{\varepsilon}(\tau) =\frac{3 \pi ^4 \Lambda ^{8/3}}{2 \tau ^{4/3}}-\frac{\pi ^2 \Lambda ^2 (4 j-3 x h)}{ 2^{7/4}\, 3^{1/4}\, \tau ^2 j^{3/4}}.
\end{equation}
Hence, using also $w^{(\Lambda)}$ to probe the hydrodynamization time will give us confidence that our analysis is not a peculiarity associated to a particular choice for $w$.

\begin{figure*}[h]
\center
\subfigure[]{\includegraphics[width=6.5cm]{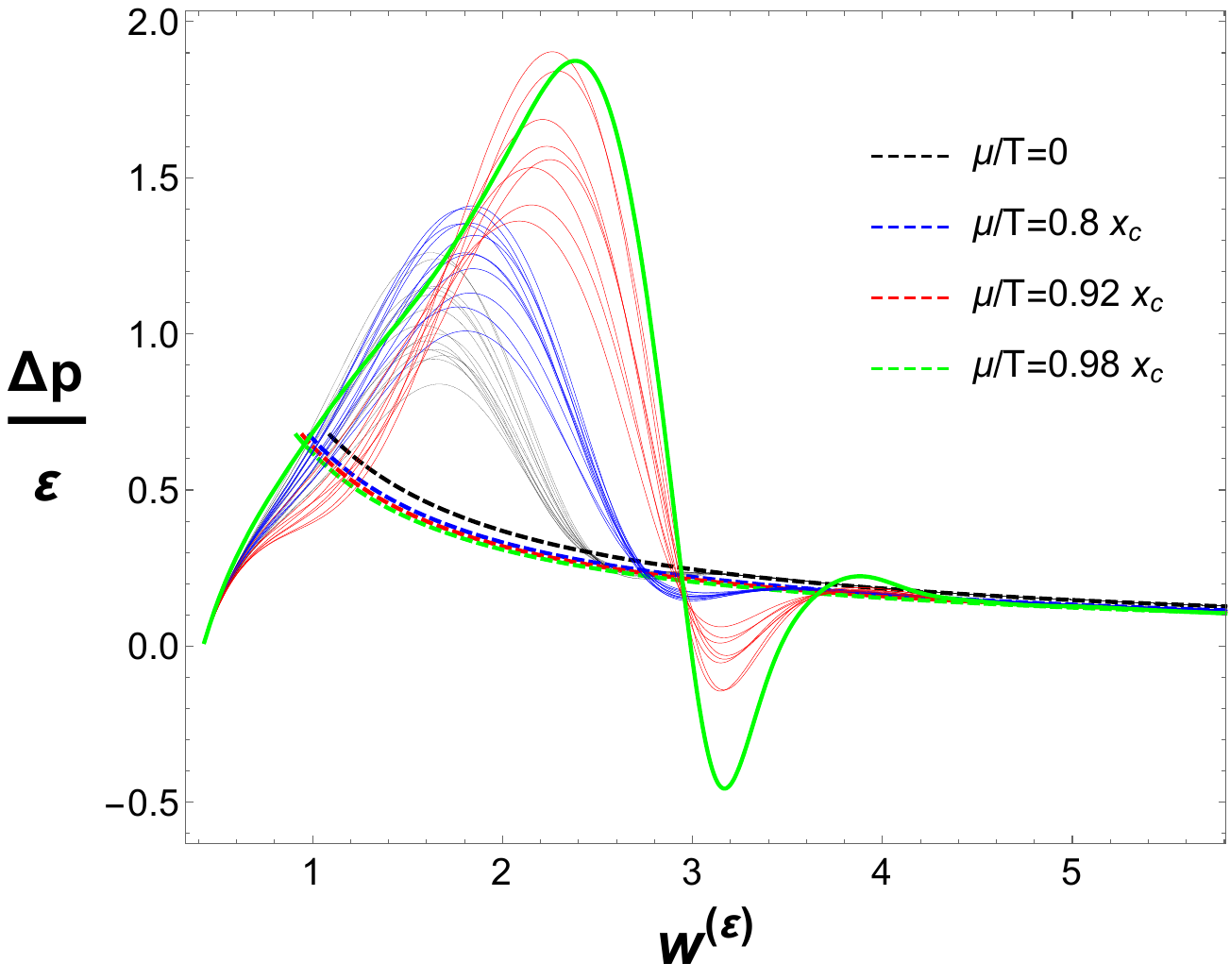}}
\qquad
\subfigure[]{\includegraphics[width=6.5cm]{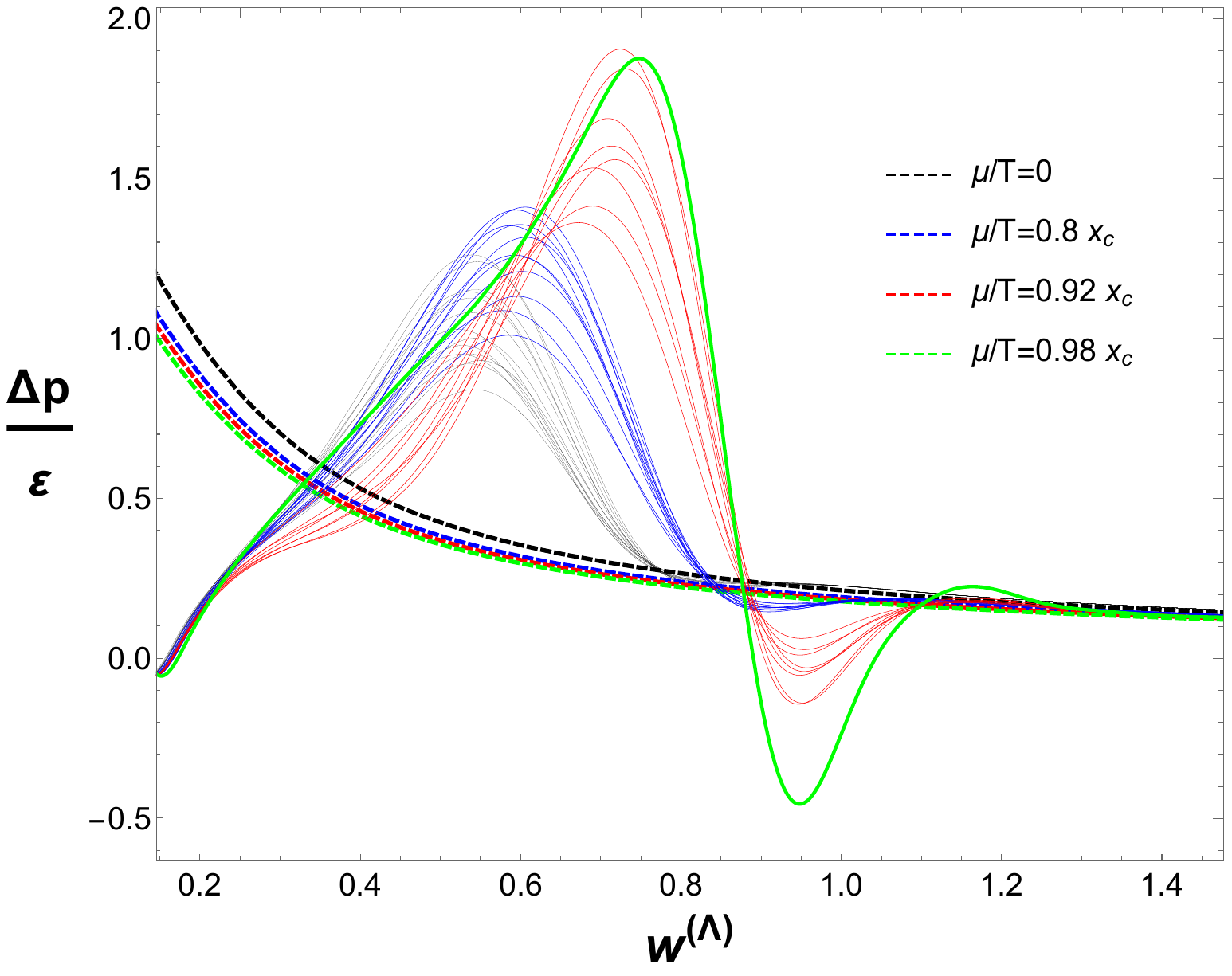}}
\qquad
\subfigure[]{\includegraphics[width=6.5cm]{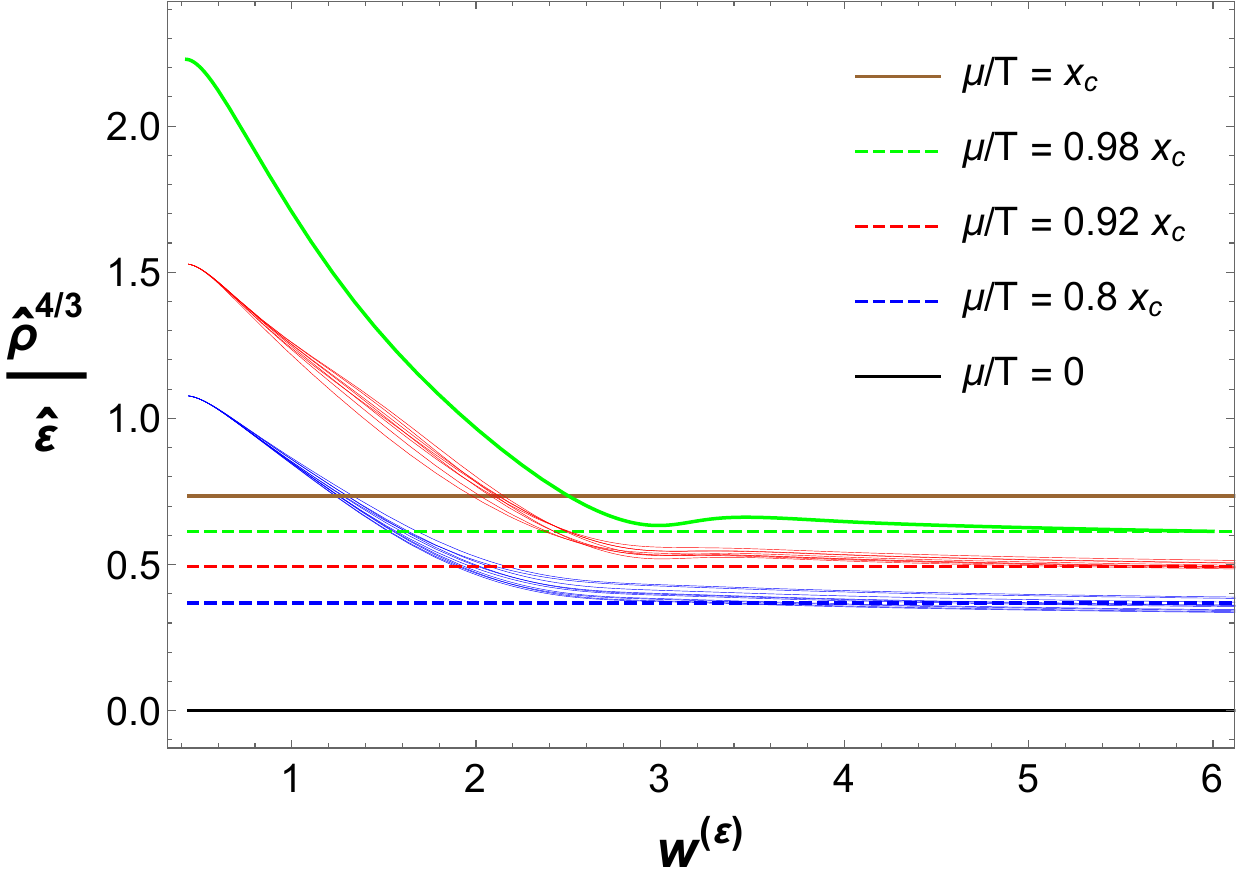}}
\qquad
\subfigure[]{\includegraphics[width=6.5cm]{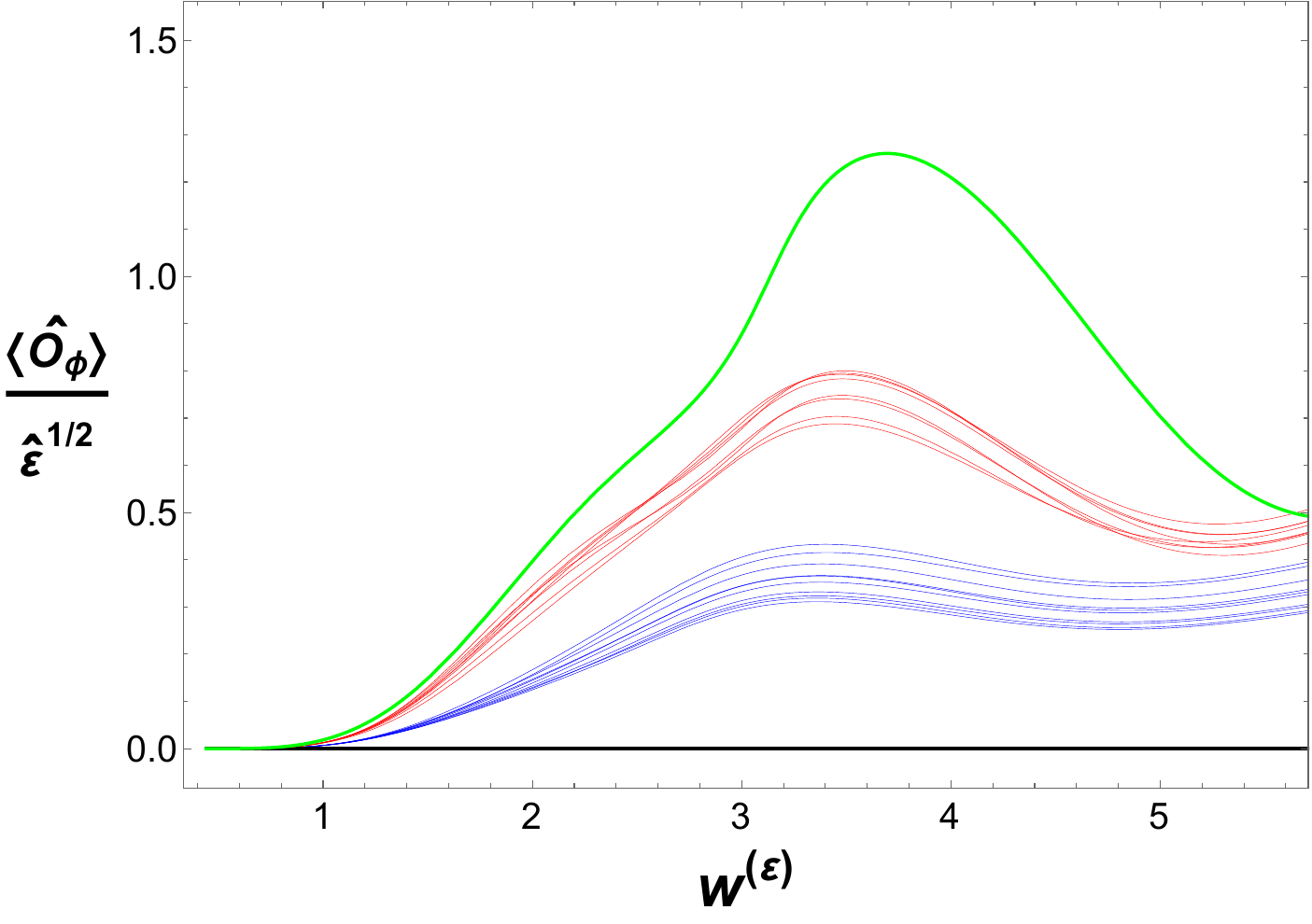}}
\caption{Holographic results for the Bjorken flow evolution of far-from-equilibrium 1RCBH backgrounds ($x_c\equiv(\mu/T)_c=\pi/\sqrt{2}$ is the critical point \cite{DeWolfe:2011ts,Finazzo:2016psx,Critelli:2017euk}). (a) Pressure anisotropy as function of $w^{(\varepsilon)}$(dashed curves are the corresponding Navier-Stokes results). (b) Pressure anisotropy as function of $w^{(\Lambda)}$ (c) Charge density. (d) Scalar condensate.}
\label{fig:result1}
\end{figure*}

\begin{figure*}[h]
\center
\subfigure[]{\includegraphics[width=7.5cm]{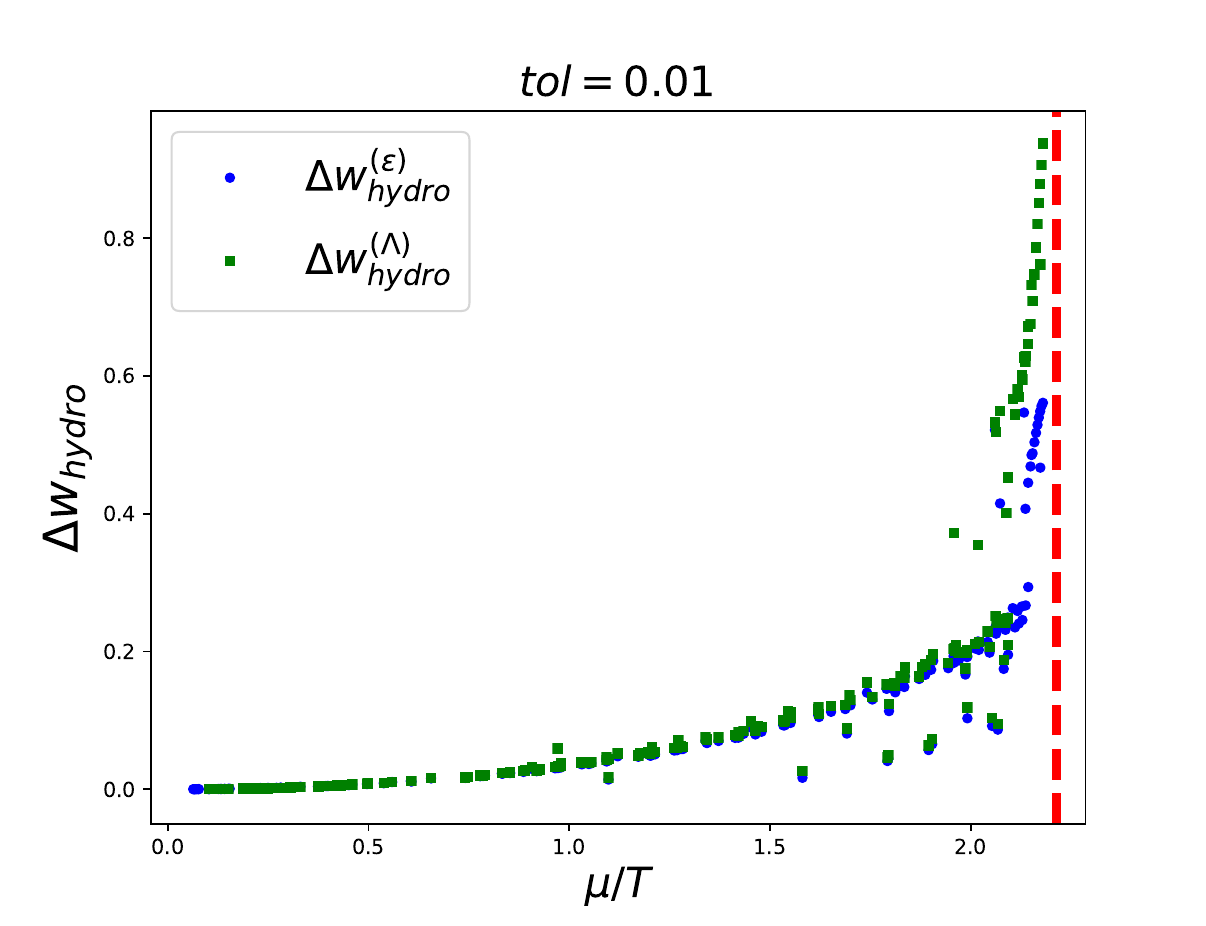}}
\qquad
\subfigure[]{\includegraphics[width=7.5cm]{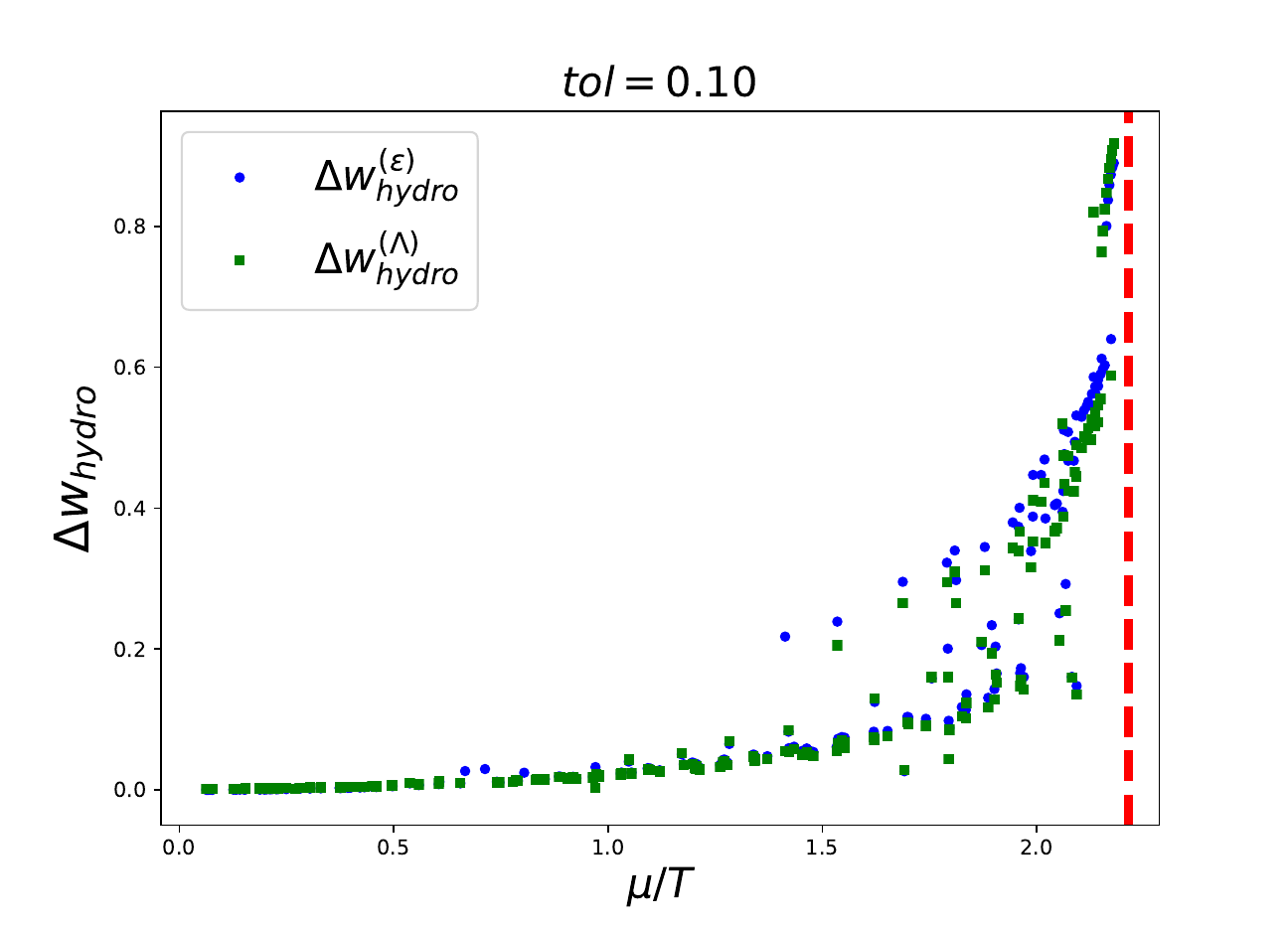}}
\caption{Variation of the hydrodynamization time with respect to the vanishing chemical potential case, as defined in Eq.\ \eqref{eq:Delta_whydro} (the vertical asymptote indicates the location of the critical point, $x_c\equiv(\mu/T)_c=\pi/\sqrt{2}$). Results using a tolerance $tol$ of (a) $1\%$ and (b) $10\%$ in Eq. \eqref{eq:hydrodynamization}.}
\label{fig:result2}
\end{figure*}

The applicability of hydrodynamics in the late time dynamics of planar shockwave collisions in SYM was first investigated in \cite{Chesler:2010bi} (see \cite{vanderSchee:2014qwa} for a comprehensive discussion of this problem). Recently, hydrodynamization was also studied in detail in non-conformal shockwave collisions in Ref.\ \cite{Attems:2017zam}. In the case of Bjorken flow, we follow the same convention of previous works \cite{Heller:2011ju,Jankowski:2014lna} and estimate the hydrodynamization time $w_{\textrm{hydro}}^{(\varepsilon/\Lambda)}$ as the timescale after which a given far-from-equilibrium numerical solution becomes well described by NS hydrodynamics, satisfying
\begin{equation}\label{eq:hydrodynamization}
\left\vert \left(\frac{\Delta p}{\varepsilon}\right)_{\textrm{numerical}} - \left(\frac{\Delta p}{\varepsilon}\right)_{\textrm{hydro}} \right\vert \leq tol \left(\frac{\Delta p}{\varepsilon}\right)_{\textrm{hydro}},
\end{equation}
where $tol$ is the tolerance threshold for the difference in the above inequality. In particular, we choose $tol=0.01$ and $tol=0.1$ to verify how robust our results are.

\section{Results}
\label{sec:Results}

Now we present our results for the far-from-equilibrium Bjorken flow in the 1RCBH model (the numerical details can be found in the Appendix). In Fig.\ \ref{fig:result1} we show the time evolution of the relevant observables characterizing the Bjorken expansion for a sample of numerical solutions with many different initial conditions (see the Appendix) and four different equilibrium values of $\mu/T$. Subfigures (a) and (b) are the most important ones, as they show the evolution of the pressure anisotropy along with the coalescence to hydrodynamics at late times, with the dashed lines indicating the analytical NS result associated with \eqref{eq:eom7}. It is clear from subfigures (a) and (b) that, as the value of $\mu/T$ increases and approaches the critical point, the numerical solution takes longer to coalesce to hydrodynamics. Subfigure (c) shows the evolution of the charge density $\hat{\rho}\equiv\kappa_5^2 \rho$, whose asymptotic behavior is used to extract the equilibrium value of $\mu/T$ in each simulation. Finally, subfigure (d) shows the evolution of the scalar condensate \eqref{eq:Ophi_Bjork_C}, which similarly to the equilibrium solution \cite{Critelli:2017euk}, increases with increasing values of the charge density.

Regarding the hydrodynamization time defined in Eq.\ \eqref{eq:hydrodynamization}, we plot in Fig.\ \ref{fig:result2} the results for its relative variation (as a function of $\mu/T$) with respect to the SYM result at vanishing chemical potential
\begin{equation}\label{eq:Delta_whydro}
\Delta w_{\textrm{hydro}}^{(\varepsilon/\Lambda)}\equiv\frac{w_{\textrm{hydro}}^{(\varepsilon/\Lambda)}(\mu/T) - w_{\textrm{hydro}}^{(\varepsilon/\Lambda)}(0)}{w_{\textrm{hydro}}^{(\varepsilon/\Lambda)}(0)}.
\end{equation}
Fig.\ \ref{fig:result2} emphasizes, in a very clear way, the main result of Fig.\ \ref{fig:result1} and of this work: the onset of hydrodynamics is significantly delayed as the chemical potential is increased towards its value at the critical point of the phase diagram, with $\Delta w_{\textrm{hydro}}^{(\Lambda)}$ being even larger than $\Delta w_{\textrm{hydro}}^{(\varepsilon)}$. Furthermore, subfigures (a) and (b) show that this qualitative effect is also robust against variations of the tolerance $tol$ in Eq.\ \eqref{eq:hydrodynamization}.

\section{Conclusion}
\label{sec:conclusion}

In the present work we presented a \emph{first principles holographic calculation} of the full nonlinear evolution of a hot and dense (i.e., $\mu\neq 0$) far-from-equilibrium strongly coupled relativistic fluid with a critical point. We investigated how the top-down holographic construction corresponding to the 1RCBH model, which describes a superconformal non-Abelian plasma with a chemical potential, evolves in space and time undergoing Bjorken flow. We found that increasing $\mu/T$ towards its critical value considerably delays the emergence of hydrodynamic behavior, as defined by the relativistic Navier-Stokes equations. This feature of the 1RCBH model, if also applicable to QCD, could imply in important differences for correlation functions calculated in and out of equilibrium, with direct impact on the experimental searches for the QCD critical point, since the main signatures of the critical point are usually considered to be the cumulants of fluctuations of conserved charges \cite{Stephanov:2011pb,Mukherjee:2015swa,Mukherjee:2016kyu}.

Regarding the far-from-equilibrium properties of the 1RCBH model, it would also be interesting to investigate the presence of hydrodynamic attractor behavior in Bjorken flow \cite{Heller:2015dha,Romatschke:2017vte,Florkowski:2017olj,Strickland:2017kux,Spalinski:2017mel,Denicol:2017lxn,Casalderrey-Solana:2017zyh} and understand how the critical point affects the properties of such an attractor. Previous studies in the case of SYM at $\mu=0$ and Gauss-Bonnet holography were already done in \cite{Spalinski:2017mel} and \cite{Casalderrey-Solana:2017zyh}, respectively.

We stress that the 1RCBH model has important differences with respect to the QGP. For instance, the 1RCBH model is conformal, while the QGP is highly non-conformal in the crossover region. Moreover, the dynamic universality classes \cite{Hohenberg:1977ym} are different: while QCD is expected to be in the type H dynamic universality class \cite{Son:2004iv}, the 1RCBH system is of type B \cite{DeWolfe:2011ts}. The fact that the dynamic universality class of the 1RCBH model is of type B is interesting because in this case $\eta/s$ is finite at the critical point, which means that viscous hydrodynamics is, in principle, well defined everywhere in the phase diagram. Nonetheless, it is also important to stress that phenomenologically realistic non-conformal holographic settings at finite temperature and chemical potential with a critical point may be constructed using the class of bottom-up holographic models first proposed in Ref.\ \cite{DeWolfe:2010he}, which was recently investigated in \cite{Critelli:2017oub} and \cite{Knaute:2017opk}. Therefore, the results presented here may be seen as the first steps towards a more realistic holographic description of the far-from-equilibrium hot and dense QGP produced in heavy ion collisions.

\begin{acknowledgments}
R.C. acknowledges financial support by the S\~{a}o Paulo Research Foundation (FAPESP) under FAPESP grant number 2016/09263-2. R.R. acknowledges financial support by Universidade do Estado do Rio de Janeiro (UERJ) and Funda\c{c}\~{a}o Carlos Chagas de Amparo \`{a} Pesquisa do Estado do Rio de Janeiro (FAPERJ). J.N. acknowledges financial support by FAPESP under grants 2015/50266-2 and 2017/05685-2 and Conselho Nacional de Desenvolvimento Cient\'{i}fico e Tecnol\'{o}gico (CNPq).
\end{acknowledgments}

\appendix
\section*{Appendix}

This Appendix is devoted to provide details on the EMD equations of motion and their numerical solutions used in the main text. Furthermore, we provide explicit expressions for the one-point functions $\left( \langle T^{\mu\nu} \rangle, \ \langle J^{\mu} \rangle , \ \langle \mathcal{O}_\phi \rangle \right)$ used to probe the Bjorken flow dynamics of the 1RCBH model. We stress that many steps followed here are similar to the ones described in Ref.\ \cite{Critelli:2017euk} which considered the case of homogeneous isotropization.

\noindent \textsl{Equations of motion.} Eq. \eqref{lineElement} comprises five independent functions, which are the dilaton $\phi(\tau,r)$ and Maxwell $\Phi(\tau,r)$ fields, besides three metric coefficients $A(\tau,r)$, $B(\tau,r)$ (the so-called metric anisotropy), and $\Sigma(\tau,r)$, providing the most general Ansatz for the EMD fields consistent with the Bjorken symmetry. The resulting EMD equations of motion for the Bjorken flow constitute a set of coupled partial differential equations (PDE's) given by
\begin{subequations}
\begin{align}
\frac{1}{6} \Sigma  \left(3 \left(B'\right)^2+\left(\phi '\right)^2\right)+\Sigma '' &=0, \label{PDE1} \\
(d_{+}\Sigma)' +\frac{2 \Sigma '}{\Sigma }d_{+}\Sigma+\frac{1}{12} \Sigma \left(f \mathcal{E}^2+2 V\right) &=0, \label{PDE2} \\
\Sigma \, (d_{+}B)'+\frac{3 \Sigma'}{2}d_{+}B+\frac{3 d_{+}\Sigma}{2} B' &=0, \label{PDE3} \\
4 \Sigma (d_{+}\phi)'+6 \phi'd_{+}\Sigma \notag \\
+6 \Sigma'd_{+}\phi+\Sigma \partial_\phi f \mathcal{E}^2-2 \Sigma  \partial_{\phi}V &=0, \label{PDE4}\\
A''  + \frac{1}{12} \left(18 B'd_{+}B-\frac{72 \Sigma'd_{+}\Sigma}{\Sigma^2} \right.\notag \\
   \left.  +  6 \phi'd_{+}\phi-7 f \mathcal{E}^2-2 V\right) & =0, \label{PDE5}\\
\frac{(\partial_{\phi}f )\phi '}{f}+\frac{3 \Sigma '}{\Sigma } + \frac{\mathcal{E}'}{\mathcal{E}} &=0 \label{PDE6}
\end{align}
\end{subequations}
where $'\equiv\partial_r$, $\mathcal{E}\equiv -\partial_r\Phi$, and $d_{+}\equiv \partial_{\tau}+A(\tau,r)\partial_r$. From Eq.\ \eqref{PDE6}, one obtains,
\begin{equation}\label{eq:Efield2}
\mathcal{E}(\tau,r) = 2\Phi_{2}(\tau)\tau\Sigma(\tau,r)^{-3}e^{2\sqrt{\frac{2}{3}}\phi(\tau,r)},
\end{equation}
where $\Phi_2$ is a time-dependent coefficient following from the near-boundary expansion of $\Phi$.

To perform the near-boundary expansion of the bulk EMD fields, which is needed to fix the boundary conditions corresponding to the Bjorken symmetry and obtain the one-point functions of the field theory undergoing Bjorken expansion at the boundary, we set as the boundary condition for the metric field the usual expression for the line element in Bjorken flow using Milne coordinates
\begin{equation}\label{eq:Bjork_line}
ds^2 = -d\tau^2 + \tau^2 d\xi^2 + dx^2+dy^2.
\end{equation}
Hence, by imposing that at the boundary the bulk metric field in Eq.\ \eqref{lineElement} is conformally equivalent to Eq.\ \eqref{eq:Bjork_line}, while the dilaton vanishes and the Maxwell field reduces to the chemical potential of the field theory, one works out the following near-boundary expansions
\begin{subequations}
\begin{align}
A(\tau,r) & = \frac{1}{2}(r+\lambda(\tau))^2-\partial_\tau\lambda(\tau) + \sum_{n=2}^{\infty} \frac{a_n(\tau)}{r^n}, \label{eq:expA}\\
\Sigma(\tau,r) & = \tau^{1/3}r+\frac{1+3 \tau  \lambda (\tau )}{3 \tau ^{2/3}}-\frac{1}{9 r \tau ^{5/3}}+\frac{5+9 \tau  \lambda (\tau)}{81 r^2 \tau ^{8/3}} \notag \\
& + \sum_{n=3}^{\infty} \frac{\sigma_n(\tau)}{r^n}, \label{eq:expSig}\\
B(\tau,r) & = -\frac{2}{3 r \tau }-\frac{2 \log (\tau )}{3}+\frac{1+2 \tau  \lambda (\tau )}{3 r^2 \tau ^2} \notag \\
 & -\frac{2+6 \tau  \lambda(\tau )+6 \tau ^2 \lambda (\tau )^2}{9 r^3 \tau ^3}+ \sum_{n=4}^{\infty} \frac{b_n(\tau)}{r^n},, \label{eq:expB} \\
\phi(\tau,r) & = \sum_{n=2}^{\infty} \frac{\phi_n(\tau)}{r^n}, \label{eq:expphi}\\
\Phi(\tau,r) & = \Phi _0(\tau ) + \sum_{n=2}^{\infty} \frac{\Phi_n(\tau)}{r^n}, \label{eq:expPhi}
\end{align}
\end{subequations}
where $\Phi_0(\tau\to\infty)=\mu$ is the gauge theory chemical potential and $\lambda(\tau)$ is an arbitrary function. This function is associated with the residual diffeomorphism invariance of the line element \eqref{lineElement} under radial shifts of the form $r\to r+\lambda(\tau)$ \cite{Chesler:2013lia}.

\noindent \textsl{One-point functions.} The relevant observables used to probe the hydrodynamization properties of the system, i.e. the energy-momentum tensor $\langle T_{\mu\nu} \rangle$, the U(1) four-current $\langle J_{\mu}\rangle$, and the scalar condensate $\langle\mathcal{O}_{\phi}\rangle$ are obtained from the one-point functions via holographic renormalization. Their formulas for the 1RCBH model are \cite{Critelli:2017euk}
\begin{align}
\kappa_{5}^{2}\langle T_{\tau\tau} \rangle & = -3a_2-\frac{1}{6}\phi_{2}^{2},\label{eq:Ttautau}\\
\kappa_{5}^{2}\langle T_{xx} \rangle & = -3a_2-\frac{1}{6}\phi_{2}^{2}-\frac{3}{2}\tau \partial_\tau a_2 -\frac{1}{6}\tau\phi_{2}\partial_\tau \phi_2,\\
\tau^{-2}\kappa_{5}^{2}\langle T_{\xi\xi} \rangle & =  3a_2+\frac{1}{6}\phi_{2}^{2}+3\tau \partial_\tau a_2 +\frac{1}{3}\tau\phi_{2}\partial_\tau \phi_2,\\
\kappa_{5}^{2}\langle J^{t} \rangle & = -\Phi_2(\tau) =  \frac{\rho_0}{\tau},\\
\kappa_{5}^{2}\langle \mathcal{O}_{\phi} \rangle & = -\phi_2, \label{eq:Ophi_Bjork_C}
\end{align}
where $\rho_0$ is an input corresponding to the initial charge density. The near-boundary coefficients $a_2$ and $\phi_2$  are dynamical quantities and they can be determined once the equations of motion are solved. The energy density, the parallel and longitudinal pressures, and the charge density are, respectively,
\begin{align}
\varepsilon \equiv \langle T_{\tau\tau} \rangle, \ \ \ p_{T} \equiv \langle T^{x}_{\ x} \rangle,\ \ \ p_{L} \equiv \langle T^{\xi}_{\ \xi}\rangle,  \ \ \ \rho \equiv \langle J^{t} \rangle.
\end{align}

\noindent \textsl{Numerics.} To solve numerically the PDE's \eqref{PDE1} --- \eqref{PDE6}, we first redefine the radial domain in such a way that the new domain is finite, i.e., $u \equiv 1/r$, which means that in this new radial coordinate the boundary lies at $u=0$.

The next step is to define subtracted fields $X_s$ by removing the trivial information encoded in the leading order terms of the near-boundary expansions of the bulk fields $X$,\footnote{Notice that $(d_{+}X)_s\neq d_{+}(X_s)$, where $X\in\lbrace \Sigma, B, \phi \rbrace$.}
\begin{align}\label{eq:Redefinitions_Bjork_C}
u^2A_{s} &= A - \frac{1}{2}\left(\frac{1}{u}+\lambda\right)^2+\partial_\tau\lambda,  \\
u^3\Sigma_{s} &= \Sigma - \tau^{1/3}\frac{1}{u} -\frac{1+3 \tau  \lambda }{3 \tau ^{2/3}}+\frac{u}{9  \tau ^{5/3}} \notag \\
 & \ -\frac{5+9 \tau  \lambda}{81 \tau ^{8/3}}u^2, \\
u^4B_{s} &= B(\tau, u) +\frac{2u}{3 \tau }+\frac{2 \log (\tau )}{3}-\frac{1+2 \tau  \lambda}{3 \tau ^2}u^2 \notag \\
& \ +\frac{2+6 \tau  \lambda+6 \tau ^2 \lambda^2}{9 \tau ^3}u^3,  \\
u^2 \phi_{s} & = \phi, \\
u^3\left(d_{+}B\right)_{s} &= d_{+}B + \frac{1}{3 \tau }-\frac{u}{3 \tau ^2}+\frac{u^2 (\tau  \lambda+1)}{3 \tau ^3}, \\
u^2 \left(d_{+}\Sigma\right)_{s} & =d_{+}\Sigma -\frac{10 u}{81 \tau ^{8/3}}-\frac{\tau^{1/3}}{2 u^2} \notag \\
 & \ -\frac{(1+\tau  \lambda) (-1+3 \tau  \lambda)}{6 \tau^{5/3}}-\frac{1+3 \tau  \lambda}{3 u \tau ^{2/3}}, \\
u \left(d_{+}\phi\right)_{s} & = d_{+}\phi, \\
\mathcal{E}_s & = \mathcal{E}.
\end{align}
We then rewrite Eqs. \eqref{PDE1} --- \eqref{PDE6} in terms of the variables $\lbrace A_s,\Sigma_s, B_s, \phi_s,(d_{+}B)_{s},(d_{+}\Sigma)_{s} ,(d_{+}\phi)_{s},\mathcal{E}_s \rbrace$. We do not write them explicitly here because the expressions are lengthy and not particularly enlightening.

Since we solve the radial problem using the pseudospectral method \cite{boyd01}, the radial $u-$grid is given by the Chebyshev-Gauss-Lobatto grid 
\begin{equation}
u_{k} = \frac{u_{\star}}{2}\left(1+\cos\left(\frac{k\pi}{N-1}\right)\right), \ \ \ k=0,\dots , N-1,
\end{equation}
where $N$ is the number of grid points, also known as collocant points. Here, $u_{\star}$ defines the infrared (IR) limit of the radial grid. The IR limit of the radial $u-$grid must be chosen in such a way that it covers the entire portion of the bulk geometry causally connected to the boundary, which means that the \textit{locus} of the event horizon is a good place to set the IR limit. Commonly, this is done by tracking the apparent horizon $u_h$,
\begin{equation}\label{eq:ap_hor}
d_{+}\Sigma \vert_{u=u_{h}} = 0,
\end{equation}
and then using $\lambda(\tau)$ to fix the position of $u_h$ in the course of the simulation. In our numerical algorithm, we approximated the above condition by using the following relation
\begin{equation}\label{eq:mod_ap_hor}
d_{+}\Sigma \vert_{u=u_\star} = \delta, 
\end{equation}
where $\delta$ is a small negative number, typically of the order $\mathcal{O}(10^{-3})$. Since $d_{+}\Sigma$ is a monotonically decreasing function for increasing values of $u$, Eq.\ \eqref{eq:mod_ap_hor} tells us that $u_{\star}$ is a little bit behind the apparent horizon $u_h$; consequently, if the radial grid spans $u \in [0,u_\star]$, we ensure that it covers the whole portion of the bulk geometry causally connected to the boundary. For the initial conditions considered in this work, which are given at the end of this Appendix, one can start with a radial grid $u \in [0,1]$ with $\lambda(\tau_0)=0$, where $\tau_0$ is the initial time used in the simulation, and in all the cases we considered, the condition \eqref{eq:mod_ap_hor} could be found after the following steps: i) one starts with $d_{+}\Sigma(\tau,u=1)$, which is generally equal to some negative number (and, therefore, we know that at this point we are behind the horizon, since at the apparent horizon Eq.\ \eqref{eq:ap_hor} holds, and beyond it $d_{+}\Sigma>0$); ii) as the simulation proceeds and we march towards the boundary, the value of $d_{+}\Sigma$ begins to increase and when the condition \eqref{eq:mod_ap_hor} is met, the corresponding value of $u_\star$ has been found; iii) in order to keep this value of $u_\star$ fixed through the numerical simulation for a given initial condition, we impose that $\partial u_\star/\partial \tau = 0$. Bearing this condition in mind, and manipulating the field equations \eqref{PDE1} --- \eqref{PDE6} at $u=u_\star$, the following relation is found (where we use that $X'(r)=-u^2 X'(u)$),
\begin{equation}\label{eq:Astar_rel}
A = 2\frac{\Sigma  \left(\left(3(d_{+}B)^2+(d_{+}\phi)^2 \right)\Sigma +6u_\star^2A' \delta\right)}{(2V+f\mathcal{E}^2)\Sigma ^2-24u_\star^2\Sigma'\delta},
\end{equation}
where all the functions in this expression are assumed to be evaluated at $u=u_\star$. In the limit where $\delta\rightarrow 0$ we recover the usual expression for the stationary apparent horizon \cite{Chesler:2013lia}. Expressing $A$ in terms of the subtracted field $A_s$ in the LHS of Eq.\ \eqref{eq:Astar_rel}, one obtains the following expression for $\partial_\tau\lambda$
\begin{align}\label{eq:dtlamb_bjor}
\partial_\tau\lambda & = u_\star^2 A_s+\frac{\lambda ^2}{2}+\frac{1}{2 u_\star^2}+\frac{\lambda }{u_\star} \notag \\
 & \ -2\frac{\Sigma  \left(\left(3(d_{+}B)^2+(d_{+}\phi)^2 \right)\Sigma +6u_\star^2A' \delta\right)}{(2V+f\mathcal{E}^2)\Sigma ^2-24u_\star^2\Sigma'\delta}.
\end{align}
Thus, the condition $\partial u_\star/\partial \tau = 0$ provides an expression for $\partial_\tau\lambda$, which is used to update the value of $\lambda(\tau)$ through the simulation keeping fixed the position of the approximated apparent horizon $u_\star$.

To solve the radial problem using the pseudospectral method, one needs the boundary conditions that are derived using the near-boundary expansion of the subtracted bulk fields
\begin{align}
\Sigma(\tau, u=0)_{s} & = \frac{-20-60\tau\lambda - 54\tau^2\lambda^2 -27\tau^4\phi_2^2}{486 \tau^{11/3}},\\
\left(d_{+}\Sigma(\tau, u=0)\right)_{s} & = a_2 \tau^{1/3}-\frac{5 (6 \lambda  \tau+5)}{243 \tau^{11/3}} \notag \\
 & \ +\frac{1}{12}\tau^{1/3} \phi_2^2,\\
\left(d_{+}B(\tau, u=0)\right)_{s} & = 2 a_2+\frac{3 \tau\partial_\tau a_2  }{2}+\frac{\lambda ^2}{3 \tau ^2}+\frac{2 \lambda }{3 \tau ^3}+\frac{1}{3 \tau ^4} \notag \\
 & \ \frac{ \tau \phi_2 \partial_\tau\phi_2}{6}+\frac{\phi_2^2}{9},\\
\left(d_{+}\phi(\tau, u=0)\right)_{s} & = -\phi_2 ,\\
A(\tau, u=0)_{s} &= a_2, \\
\partial_u A(\tau, u=0)_{s} &= \frac{\partial_\tau a_2}{2}-2\lambda a_2.
\end{align}

Thus, in order to solve the radial problem at a given time $\tau$, one has to know\footnote{If we know $B(\tau,u)$ and $a_2(\tau)$, then we know $\partial_\tau a_2$. This is because the term $b_4(\tau)$ is proportional to $a_2(\tau)$ and $\partial_\tau a_2$:
\begin{align}\label{eq:b4_bjorC}
b_4(\tau)& = B(\tau, u=0)_{s} = -a_2(\tau )-\frac{3}{4} \tau  \partial_\tau a_2(\tau )+\frac{1}{6 \tau ^4} \notag \\
 & \ +\frac{2 \lambda (\tau )}{3 \tau ^3}+\frac{\lambda (\tau )^2}{\tau^2}+\frac{2 \lambda (\tau )^3}{3 \tau } - \frac{1}{18}\phi_2(\tau)^2-\frac{1}{12}\phi_2(\tau)\partial_\tau\phi_2(\tau).
\end{align}} $\lbrace B(\tau,u),\phi(\tau,u), a_2(\tau), \Phi_2(\tau), \lambda(\tau) \rbrace$. Evidently, the initial state that we define must contemplate this set as well.

To perform the time evolution of the system one has to use $d_{+}\equiv \partial_{\tau}+A(\tau,r)\partial_r$ to extract $\lbrace\partial_\tau B_s, \partial_\tau \phi_s \rbrace$ from $\lbrace (d_{+} B)_s , (d_{+} \phi)_s \rbrace$; $\partial_\tau a_2$ from the coefficient $b_4(\tau)$ of the expansion \eqref{eq:expB}; and $\partial_\tau\lambda$ from Eq.\ \eqref{eq:dtlamb_bjor}. By doing so, one can evolve in time the fields necessary to start the cascade solution of the nested set of PDE's
\begin{align}
B(\tau+\Delta\tau, u)_s & = B(\tau, u)_s + \Delta\tau  (\partial_\tau B_s ), \\
\phi(\tau+\Delta\tau, u)_s & = \phi(\tau, u)_s + \Delta\tau  (\partial_\tau \phi_s ), \\
a_2(\tau+\Delta\tau)& = a_2(\tau) + \Delta \tau (\partial_\tau a_2 ), \\
\lambda(\tau+\Delta\tau) & = \lambda(\tau) + \Delta\tau  (\partial_\tau \lambda ),
\end{align}
where $\Delta\tau$ is the time step. There are several ways to do the time evolution. In this work, we choose a fourth-order Adams-Bashforth method to evolve in time.

\noindent \textsl{Initial conditions.} Regarding the initial conditions used in the present work, they were chosen as follows
\begin{align}
B(\tau_0, u)_s & = f(u) +\frac{\alpha}{u^4} \left[-\frac{2}{3}  \log \left(u+\tau _0\right) \right. \notag \\
 & \left. -\left(-\frac{2}{3}  \log \left(\tau _0\right)-\frac{2 u^3}{9 \tau_0^3}+\frac{u^2}{3 \tau _0^2}-\frac{2 u}{3 \tau _0}\right) \right], \label{eq:Bini_bjork} \\
a_2(\tau_0) & = -20/3, \\
\phi(\tau_0, u)_s & = 0,\\
\Phi_2(\tau_0) & = -\frac{\rho_0}{\tau_0},\\
\lambda(\tau_0) & = 0, \\
\tau_0 & = 0.2,
\end{align}
where
\begin{equation}
f(u) = \sum_{i=0}^{5}\beta_i u^i.
\end{equation}
Thus, different values for the set $\lbrace \alpha, \beta_i , \rho_0 \rbrace$ will produce different evolutions and, consequently, different results for the hydrodynamization time \eqref{eq:hydrodynamization}. In particular, to generate the results presented in this paper, we have randomly selected values for $\lbrace \alpha, \beta_i , \rho_0 \rbrace$ in the range,
\begin{equation}
1\leq \alpha \leq 1.05 , \ \ \ -0.5 \leq \beta_i \leq 0.5 \ \ \text{and} \ \ 0\leq \rho_0 \lesssim 3.7.
\end{equation}
The fact  that $\rho_0$ does not have a clear upper bound, i.e. $\rho_0 \lesssim 3.7$, is because some initial conditions do not have a well-behaved evolution for e.g. $\rho_0=3.7$. The reason is because some initial profiles tend to reach the vicinity of the critical point with a smaller value of $\rho_0$. Indeed, one cannot predict the final value of $\mu/T$ by just looking at the initial conditions since this is an equilibrium quantity which is only determined by the late time evolution of the system, constituting, therefore, \textit{a posteriori} analysis of the numerical data.

Regarding our first definition of the dimensionless time flow, $w^{(\varepsilon)}\equiv\hat{\varepsilon}(\tau)^{1/4}\tau$, which was used to probe the hydrodynamization time and employed in the plots of the main text instead of an alternative definition such as $w\equiv T(\tau)\tau$, this is justified based on the following facts. First, the energy density is a physically well defined observable at all times along the evolution of the system while the temperature is a quantity which rigorously only makes sense in or at least close to thermal equilibrium. Therefore, one should track the far-from-equilibrium time evolution of the system by using an observable which is physically well defined regardless whether the system is in equilibrium or not and this is the case for the energy density, but not for the temperature. Second, if one insists in defining $w\equiv T(\tau)\tau$ in the present setup, one immediately finds an ambiguity in such definition. In fact, by looking at the \textit{equilibrium} equation of state given by Eqs.\ \eqref{eq:Temp_Bjork_C} and \eqref{eq:ahmoleke}, one can extract the relations $T=j^{-1/4}\hat{\varepsilon}^{1/4}$ and $T=h^{-1/3}\hat{\rho}^{1/3}$, which agree with each other in equilibrium. By naively trying to define and ``out-of-equilibrium temperature'' $T(\tau)$ by extrapolating these equilibrium relations to the far-from-equilibrium regime, $T(\tau)=j^{-1/4}\hat{\varepsilon}^{1/4}(\tau)$ and $T(\tau)=h^{-1/3}\hat{\rho}^{1/3}(\tau)$, one is left with an ambiguity in the definition of such ``out-of-equilibrium temperature'', since these two expressions are generally different from each other and only agree in equilibrium. Such an ambiguity is a clear manifestation of the fact that one should not use equilibrium relations when the system is far-from-equilibrium. Nonetheless, the simulations do coalesce to a near equilibrium state and the ideal fluid limit is recovered for asymptotic times $\tau\rightarrow\infty$, in which the temperature has a well defined profile in the Bjorken flow $T_{\textrm{asym}}(\tau)=\Lambda/\left(\Lambda \tau \right)^{1/3}$ \cite{Chesler:2009cy}, where $\Lambda$ is an energy scale that depends on the initial conditions. Therefore, although it is not possible to define $T(\tau)$ throughout the whole evolution, one can analyze the near equilibrium stage of the evolution to define a second dimensionless time flow, i.e. $w^{(\Lambda)}\equiv T_{\textrm{asym}}(\tau)\tau =\left(\Lambda \tau \right)^{2/3}$.

Furthermore, just as in the homogeneous isotropization case \cite{Critelli:2017euk}, we also had to use a filter to avoid spurious growth of hard modes \cite{boyd01}. This is done by accessing the spectral coefficients $\lbrace a_i \rbrace$ and damping the high frequency modes using an exponential map of the type $a_i\rightarrow e^{-a(i/N)^b}a_i$. Typical values for the pair $(a,b)$ used in our work are in the ballpark of $(16,18)$.

To check the numerical consistency of our results, besides standard tests such as the variation of number of collocant points $N$ and the size of the time step $\Delta\tau$, we have also analyzed the late time convergence of the numerical solutions to the analytical NS solution \eqref{eq:eom7}; we were also able to reproduce the results of Wilke van der Schee's code\footnote{\url{https://sites.google.com/site/wilkevanderschee/ads-numerics}} valid for $\mu/T=0$. Another non-trivial consistency check is the observation that the value of $\delta$ as defined in Eq.\ \eqref{eq:mod_ap_hor} does not change in the course of the simulation. Our code that solves the Bjorken expanding far-from-equilibrium 1RCBH model was initially developed in Mathematica, and later translated to C++ using the Eigen linear algebra package \cite{eigen} and OpenMP to parallelize the code.



%

\end{document}